\documentclass[a4paper]{article}

\pdfoutput=1

\usepackage{icrc2013}
\usepackage{subfig}
\usepackage[english]{babel}

\title{CRPropa 3.0 -- a Public Framework for Propagating UHE Cosmic Rays through Galactic and Extragalactic Space}

\newcommand{\etal}{\MakeLowercase{\textit{et al. }}} 
\shorttitle{G.\ M\"{u}ller \etal CRPropa 3.0}

\authors{Rafael Alves Batista$^{1}$, Martin Erdmann$^{2}$, Carmelo Evoli$^{1}$, Karl-Heinz Kampert$^{3}$, Daniel Kuempel$^{2}$, Gero M\"{u}ller$^{2}$, Peter Schiffer$^{1}$, Guenter Sigl$^{1}$, Arjen van Vliet$^{1}$, David Walz$^{2}$, Tobias Winchen$^{2}$}

\afiliations{$^{1}$ University of Hamburg, II Institut f\"ur Theoretische Physik Luruper Chaussee 149, 22761 Hamburg, Germany\\ $^{2}$ RWTH Aachen University, Physikalisches Institut IIIa Otto-Blumenthal-Str., 52056 Aachen, Germany\\ $^{3}$ University of Wuppertal, Department of Physics, Gau\ss str.\ 20, 42097 Wuppertal, Germany}

\email{crpropa@desy.de}

\abstract{The interpretation of experimental data of ultra-high energy cosmic rays (UHECRs) above $10^{17}$~eV is still under controversial debate. The development and improvement of numerical tools to propagate UHECRs in galactic and extragalactic space is a crucial ingredient to interpret data and to draw conclusions on astrophysical parameters. In this contribution the next major release of the publicly available code CRPropa (3.0) is presented. It reflects a complete redesign of the code structure to facilitate high performance computing and comprises new physical features such as an interface for galactic propagation using lensing techniques and inclusion of cosmological effects in a three-dimensional environment. The performance is benchmarked and first applications are presented.}


\begin{document}
\maketitle

\section{Introduction}
Recent years have seen interesting results on spectrum~\cite{Abraham:2010mj,AbuZayyad:2012ru}, composition~\cite{Abraham:2010yv,Abbasi:2009nf}, and anisotropy~\cite{Abreu:2010ab,AbuZayyad:2012hv} of ultra-high energy cosmic rays above $10^{17}\,$eV, both from the Pierre Auger Observatory and other experiments such as the Telescope Array and the High Resolution Fly's Eye (HiRes). In order to interpret these data in the context of concrete astrophysical scenarios for the distribution of the sources, their injection characteristics such as spectrum, maximal energy and mass composition, as well as for the distribution of large scale cosmic magnetic fields requires a comprehensive numerical tool that can simulate the deflection of UHECRs over several orders of magnitude in energy and length scales, ranging from hundreds of megaparsecs down to galactic scales of the order of kiloparsecs, including their interactions such as photo-disintegration, pion production and pair production. Such a tool should be highly modular, since constraining the origin of UHECRs requires simulations predicting spectra, compositions and anisotropies for a large number of astrophysical scenarios, and comparison with experimental data. To this end, CRPropa 3.0 was developed which is based on the original CRPropa 2.0~\cite{Kampert:2012fi}. In the present contribution we will summarize its main features (Sects. 2 and 4), the code structure (Sect. 3), as well as applications and a benchmark scenario (Sects. 5 and 6). \section{Inherited features from CRPropa 2.0}
CRPropa~\cite{Kampert:2012fi} is a publicly available software package designed to simulate the extragalactic propagation of UHECRs and their secondaries. The interactions that are implemented in CRPropa 2.0 are photo-disintegration, pion production and pair production on both the cosmic microwave background (CMB) and the cosmic infrared background (IRB), as well as the decay of nuclei. Furthermore, the possibility to track secondary $\gamma$-rays and neutrinos is provided. The simulations can be done either in a one-dimensional (1D) or three-dimensional (3D) mode. In the 3D mode it is possible to define a 3D source distribution and take into account the deflections of UHECRs in extragalactic magnetic fields. In contrast, in the 1D mode, cosmological and source evolution with redshift, as well as a redshift scaling of the background light intensity, can be implemented. All these features have been inherited by CRPropa 3.0. \section{Code structure and steering}
After the release of CRPropa 2.0 multiple new applications were found and new features were developed.
To account for new use cases, in CRPropa 3.0 the propagation of cosmic rays is now composed of modules which access and modify a cosmic ray candidate.
The modular structure allows to easily add new features and to use and verify all parts of the software individually. \\
\textbf{Cosmic Ray Candidate and Modules:} 
The interface between the modules is the cosmic ray candidate class.
Cosmic ray candidates contain information about all aspects of their propagation:
the particle states at different times, module specific data, the list of secondary candidates, a list of states for stochastic interactions and a list of arbitrary properties.
All information about the propagation state, including the states of the modules, is stored in the candidates themselves.
This way modules can concurrently process multiple candidates, which is required for high performance parallel computing.
Cosmic ray candidates can be created manually or by a modular source model class, which is composed of multiple source properties, e.g.\ position, spectrum and composition.

 \begin{figure}[t]
 \centering
 \includegraphics[width=8cm]{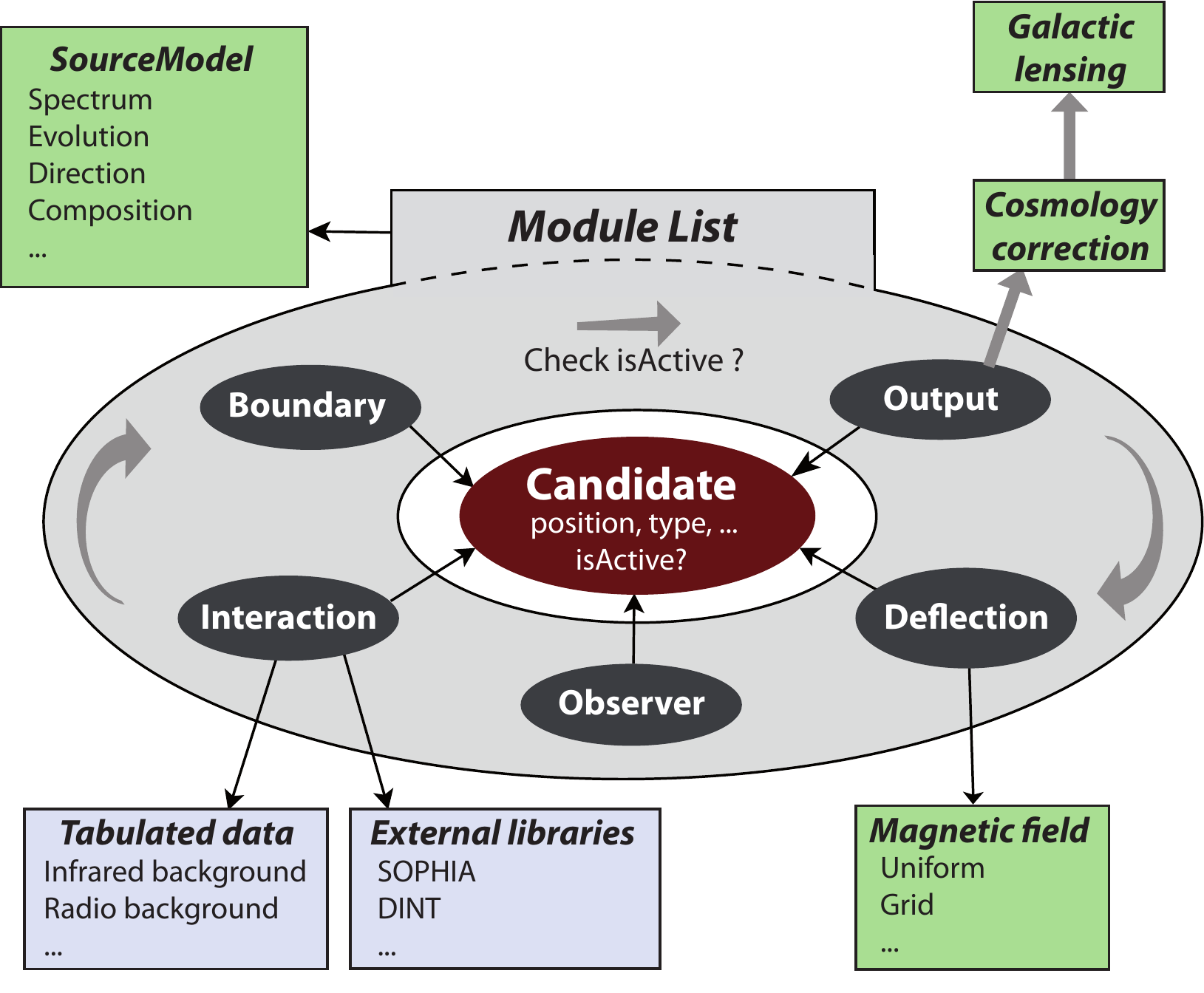}
 \caption{
   Illustration of the CRPropa 3.0 modular structure.
   Each module contained in the module list acts on the candidate class.
   The \texttt{IsActive} flag serves as break condition and is checked after each cycle of the module list.
   }
 \label{fig:CRPropaStructure}
 \end{figure}

A graphical illustration of the propagation process is given in Fig.\ \ref{fig:CRPropaStructure}.
First cosmic ray candidates are created by the source class.
Then the modules sequentially process the cosmic ray candidate.
Typically the first module is the integration module which deflects the particle in magnetic fields and updates the position, direction and trajectory length.
After that, interaction modules change the candidate's energy and nature, usually by producing secondaries.
Stochastic interaction modules decide independently if an interaction occurs during the current step.
When an interaction occurred the states of all interactions are reset.
Boundary and observer modules check if the cosmic ray candidate is still to be considered for further propagation or deactivated, respectively.
Output modules finally store trajectory points per step, on detection or under other conditions.\\

\textbf{XML and Python Steering:}
A convenient way to use the new modular version is to use CRPropa 2.0 compatible XML steering files.
The CRPropa 3.0 executable instantiates a list of modules which mimic the behavior of CRPropa 2.0.
A more flexible way of using CRPropa 3.0 is to use its Python bindings to configure and run simulations.
All classes and modules are available in Python allowing a programmatic setup of magnetic fields, sources and modules.
It is furthermore possible to run simulations interactively or write custom modules from interactive Python shells.
\section{New features}

CRPropa 3.0 introduces several new features which are summarized below.\\

\textbf{Cosmology in 3D:}
Cosmological effects such as the redshift evolution of the photon backgrounds and the adiabatic expansion of the universe are important when simulating the propagation of UHECRs. These effects can easily be taken into account in 1D simulations. However, in 3D, when deflections due to the pervasive cosmic magnetic fields are considered, it is not possible to know {\it a priori} the effective propagation length, and therefore the redshift, of the simulated particles. The obvious solution is to perform a four-dimensional simulation, which might be time consuming and not affordable depending on the desired statistics. A possible solution to this is to introduce an {\it a posteriori} correction to account for cosmological effects.

Let $E_i$ be the initial energy of the simulated cosmic ray,  $(A_i^{\rm 3D},Z_i^{\rm 3D})$ the initial mass number and atomic number, $E_f^{\rm 3D}$ the observed energy, $(A_f^{\rm 3D},Z_f^{\rm 3D})$ the observed particle type, $T^{\rm 3D}$ the effective trajectory length, and $(\theta,\phi)$ the arrival direction. Then, by resimulating each of these observed particles in 1D, using as input the parameters $E_i^{\rm 3D}$, $(A_i^{\rm 3D},Z_i^{\rm 3D})$ and $T^{\rm 3D}$, a set of subproducts of this initial particle, arisen from the photodisintegration of the primary nucleus, can be obtained. These subproducts, indexed by $k$, are observed with energy $E_{f,k}^{\rm 1D}$, type $(A_{f,k}^{\rm 1D},Z_{f,k}^{\rm 1D})$ and with the same initial properties as the injected particles from the original 3D simulation. As an approximation we can randomly choose one of the particles indexed by $k$, and substitute it in the 3D simulation by setting $E_f^{\rm 3D}=E_{f,k}^{\rm 1D}$ and $(A_f^{\rm 3D},Z_f^{\rm 3D})=(A_{f,k}^{\rm 1D},Z_{f,k}^{\rm 1D})$.

\begin{figure}
	\includegraphics[width=1.0\columnwidth]{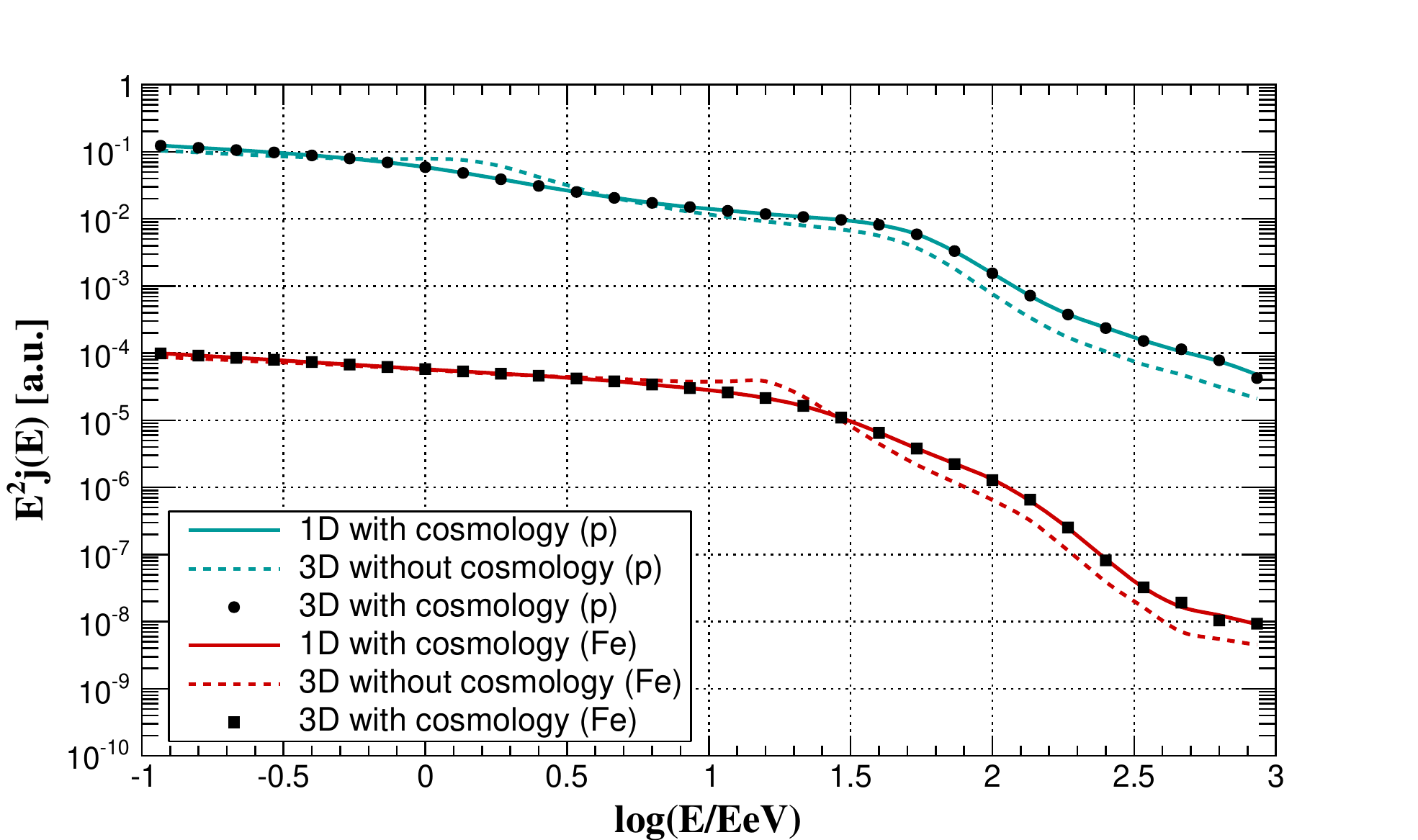}
	\caption{Comparison between the 1D spectrum (solid line), with the 3D spectrum without cosmology (dashed line) and the 3D spectrum corrected for cosmology (points), for the pure proton (cyan) and pure iron (red) cases. The $y$-axis is in arbitrary units and the iron spectra were scaled down by a factor of 1000 with respect to the proton ones. The displayed spectra correspond to simulations of a uniform source distribution up to 4000~Mpc with a differential spectral index of $-2.2$.}
	\label{fig:cosmology}
\end{figure}

The immediate test for this correction is to apply it to a 3D simulation without magnetic fields, and compare it to a 1D simulation. In this case, the only expected difference is the presence of cosmology. This comparison yields, as expected, excellent results, as shown in Fig.\ \ref{fig:cosmology}. In this figure there is a clear discrepancy between the 3D without cosmology and the 1D spectra, showing the importance of accounting for cosmological effects when simulating the propagation of UHECRs.

The error introduced in the arrival directions due to neglecting the continuous energy loss related to the adiabatic expansion of the universe is negligible for the typical strength of the extragalactic magnetic fields below $\sim \mu {\rm G}$.\\
\textbf{Galactic propagation:}
The Galactic magnetic field (GMF) is expected to significantly contribute to the total deflections of charged extragalactic UHECRs. Therefore, the functionality of CRPropa 3.0 was extended to allow forward- and backtracking of UHECRs through different models of the GMF available in the software.
Arbitrary field models can be defined using one of the grid techniques, described in the next section.
Additionally, several models in analytical form are available, including the JF12 model with both regular and random component \cite{Jansson:2012pc, Jansson:2012rt}.\\
A different, highly efficient, way to model galactic propagation is the lensing technique described and implemented in the PARSEC software \cite{Bretz:2013oka}. In this approach UHECR interactions with photons and interstellar matter are neglected due to the short distance inside the Galaxy compared to extragalactic distances.
The lensing technique uses a set of transformation matrices for different energies to map the directions of UHECRs at the border of the galaxy to directions observed at Earth.
CRPropa 3.0 provides an interface to PARSEC to apply the lensing technique on UHECRs that were propagated from an extragalactic source to the border of the Galaxy.
This combination allows to simulate the propagation of UHECRs through both the extragalactic and galactic magnetic field, which would be computationally unfeasible with pure forward tracking.\\
\textbf{Magnetic field techniques:}
Results from large scale structure simulations can be stored as smooth particles, containing the magnetic field and mass density \cite{Dolag:2008ya}.
Smooth particles have the advantage of a dynamic resolution, but lack the performance for a fast lookup of magnetic field or mass density values.
By using a third party library CRPropa 3.0 is able to use smooth particles for full resolution fields and multi-resolution grids for high performance access to precomputed fields.
\section{Applications}
As an application we compare spectrum, deflection angle and arrival directions for primary protons and iron nuclei.

In our example scenario we assume an infinite source density following the large scale structure of baryonic matter. For the baryonic matter distribution and the extragalactic magnetic field (EGMF) the large scale structure simulation of Miniati et al.\ \cite{Sigl:2004} is used. For the position of the observer, an earth-like position in the simulation box is chosen. For comparison the universal spectrum for a uniform source distribution has as well been simulated. To obtain a sufficient resolution at high energies an injection spectrum of $E^{-1}$ is simulated which is reweighted afterwards to $E^{-2.5}$. The maximum rigidity is set to $1000$~EeV.

\begin{figure*}[ht]
    \centering
    \subfloat[Spectrum]{
        \includegraphics[width=0.49\textwidth]{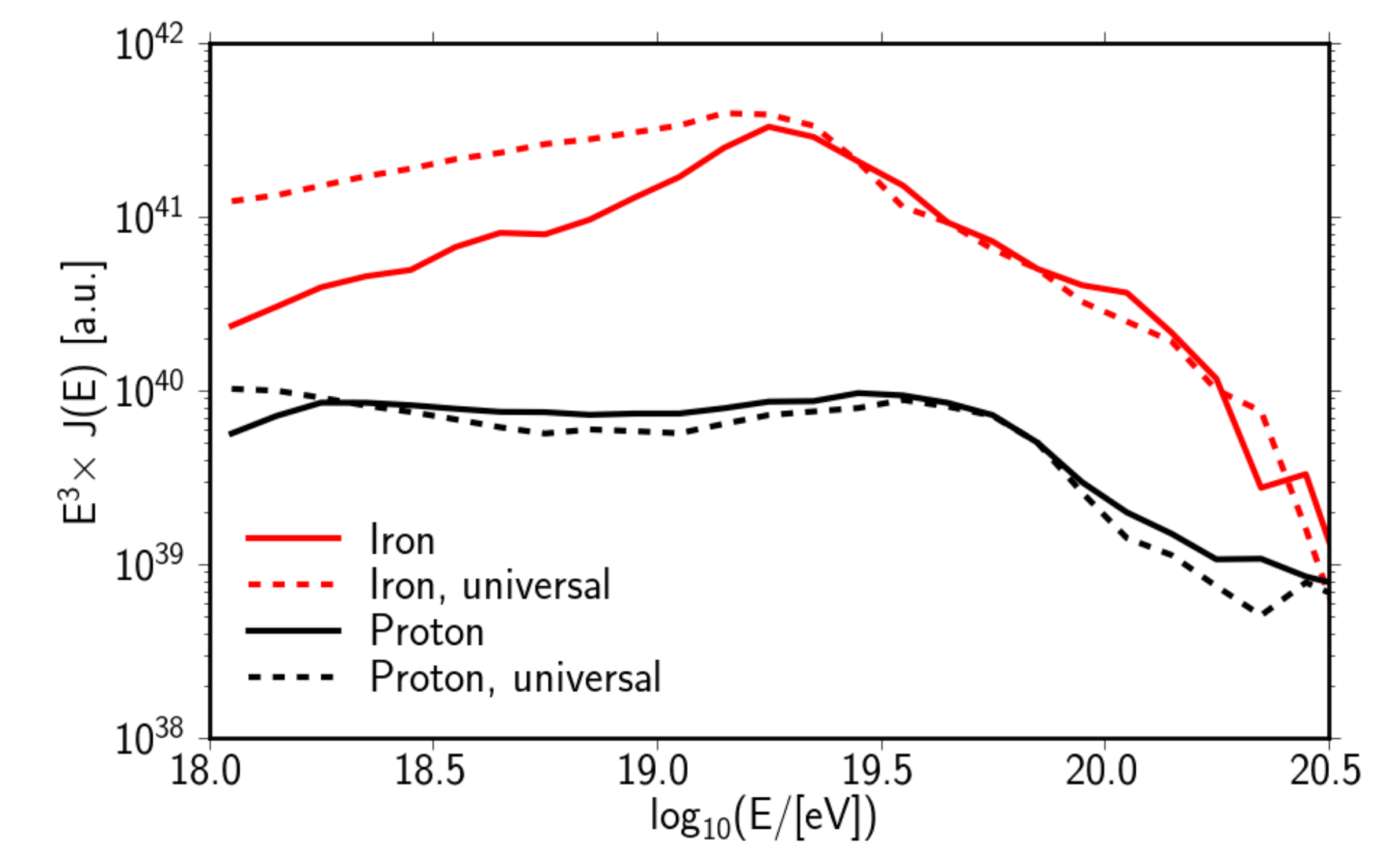}
        \label{fig:spec}
    }
    \subfloat[Mean deflection of UHECRs]{
        \includegraphics[width=0.49\textwidth]{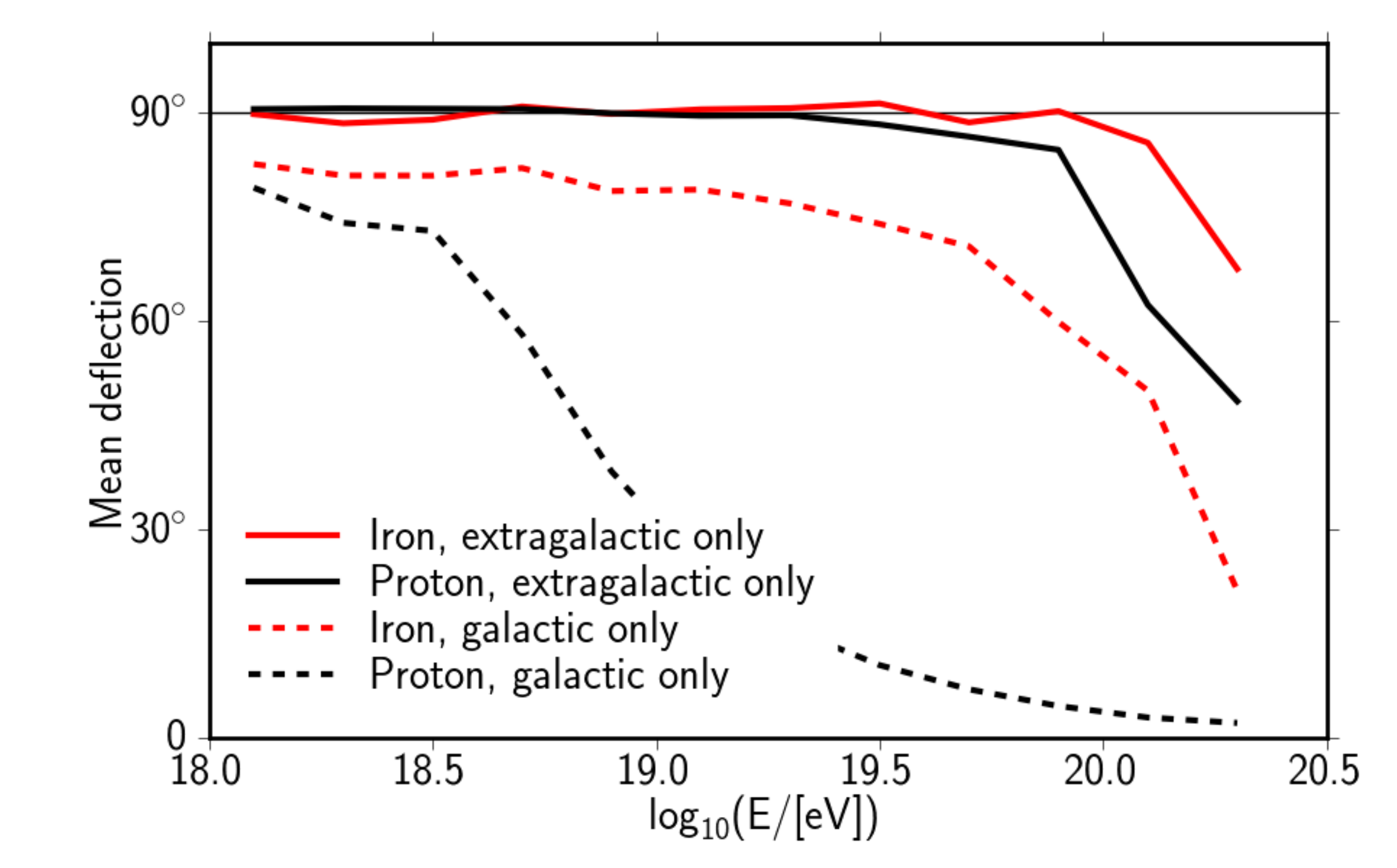}
        \label{fig:defl}
    }
    \qquad
    \subfloat[Arrival directions of injected protons at the Galactic border]{
        \includegraphics[width=0.49\textwidth]{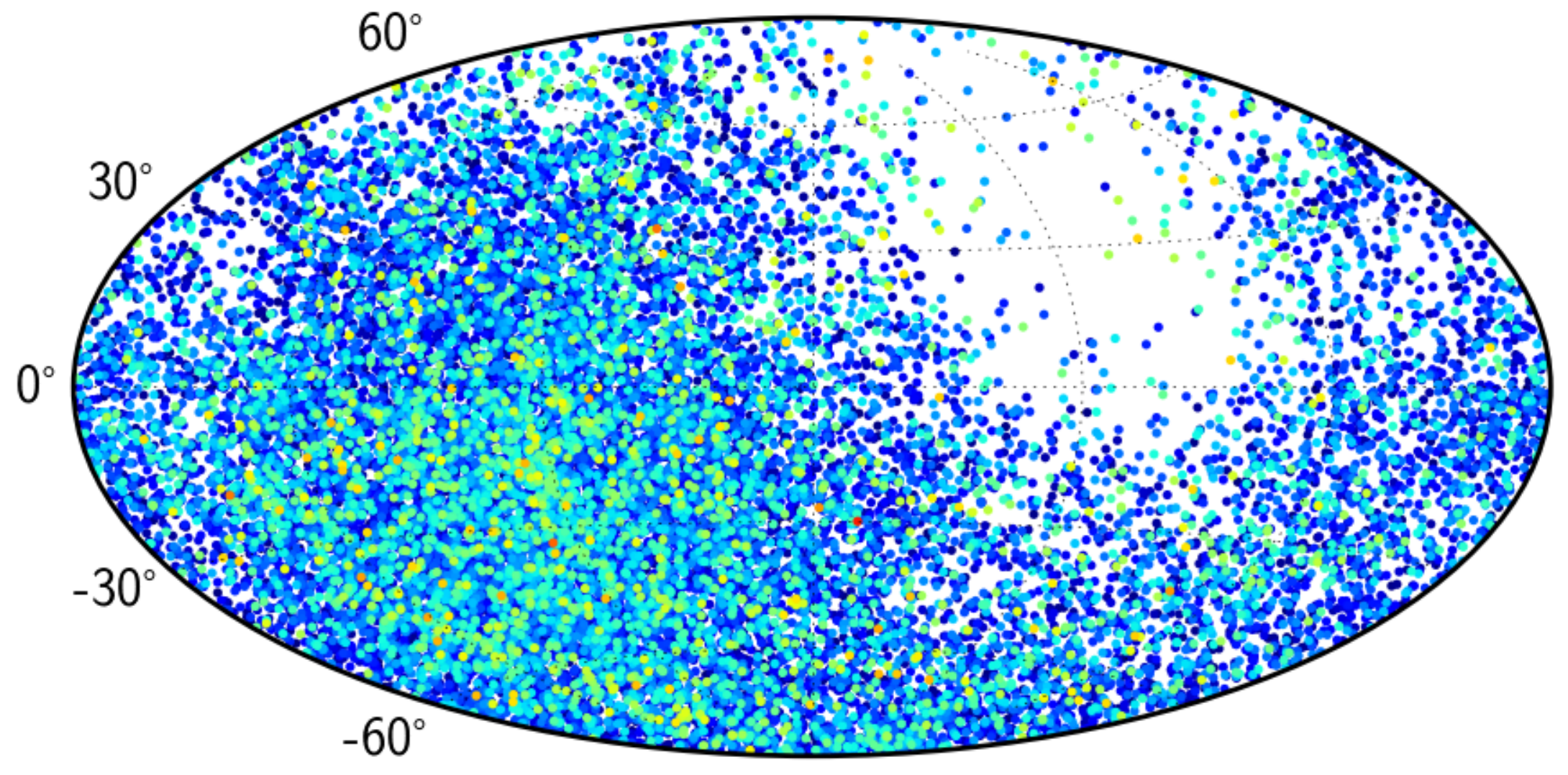}
        \label{fig:proton_xgal}
    }
    \subfloat[Arrival directions of injected iron at the Galactic border]{
        \includegraphics[width=0.49\textwidth]{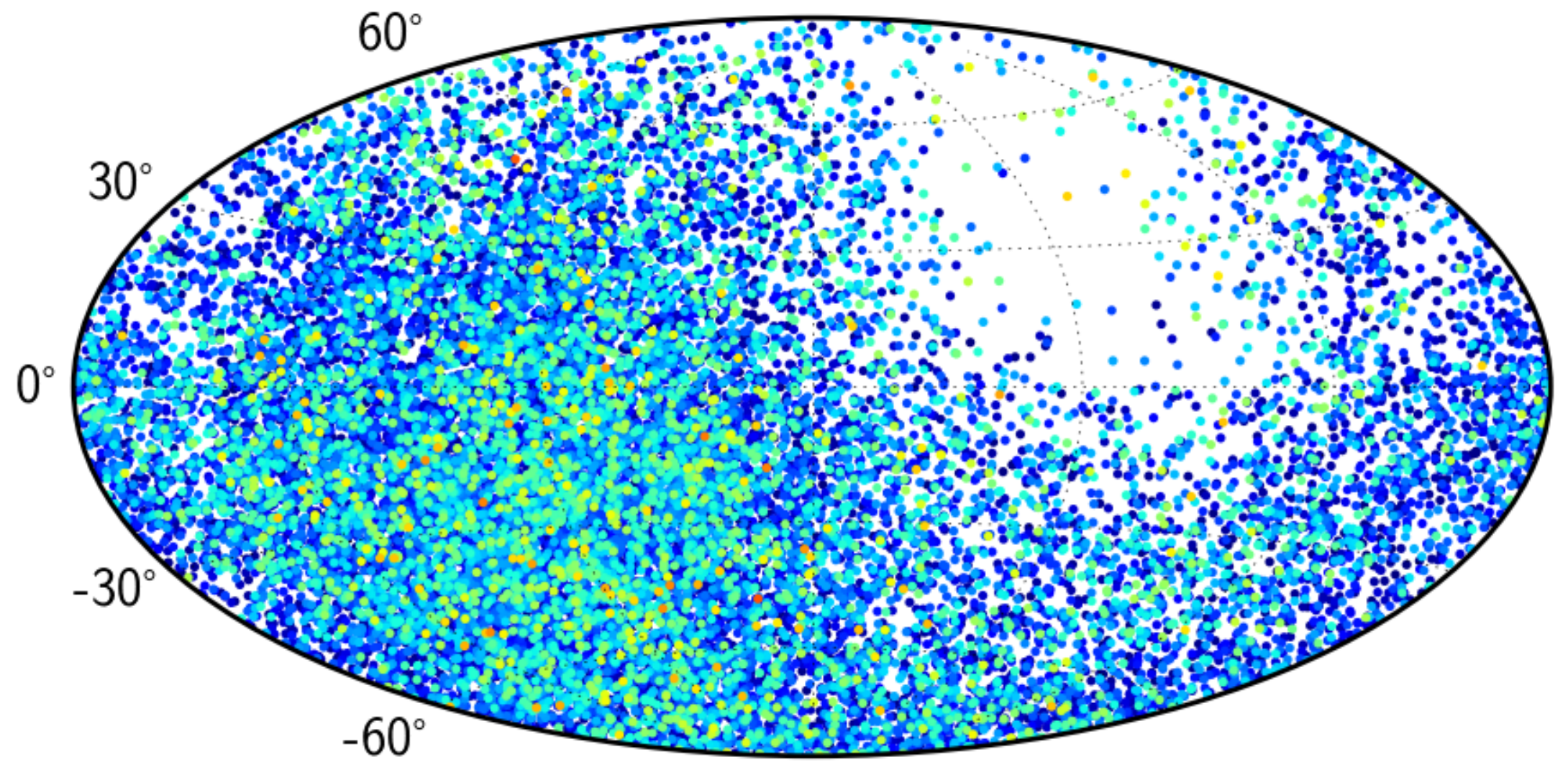}
        \label{fig:iron_xgal}
    }
    \qquad
    \subfloat[Arrival directions of injected protons at Earth]{
        \includegraphics[width=0.49\textwidth]{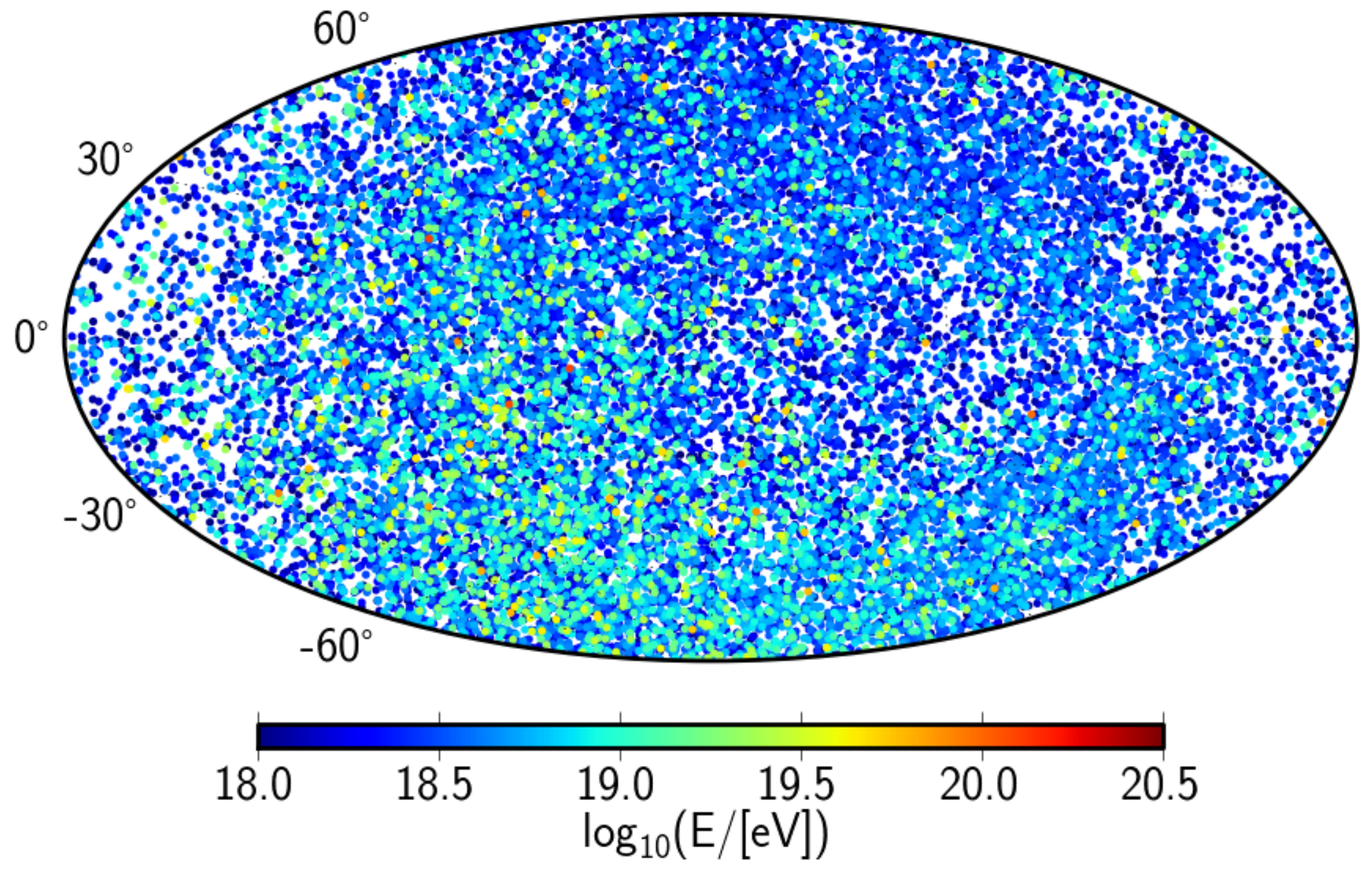}
        \label{fig:proton_gal}
    }
    \subfloat[Arrival directions of injected iron at Earth]{
        \includegraphics[width=0.49\textwidth]{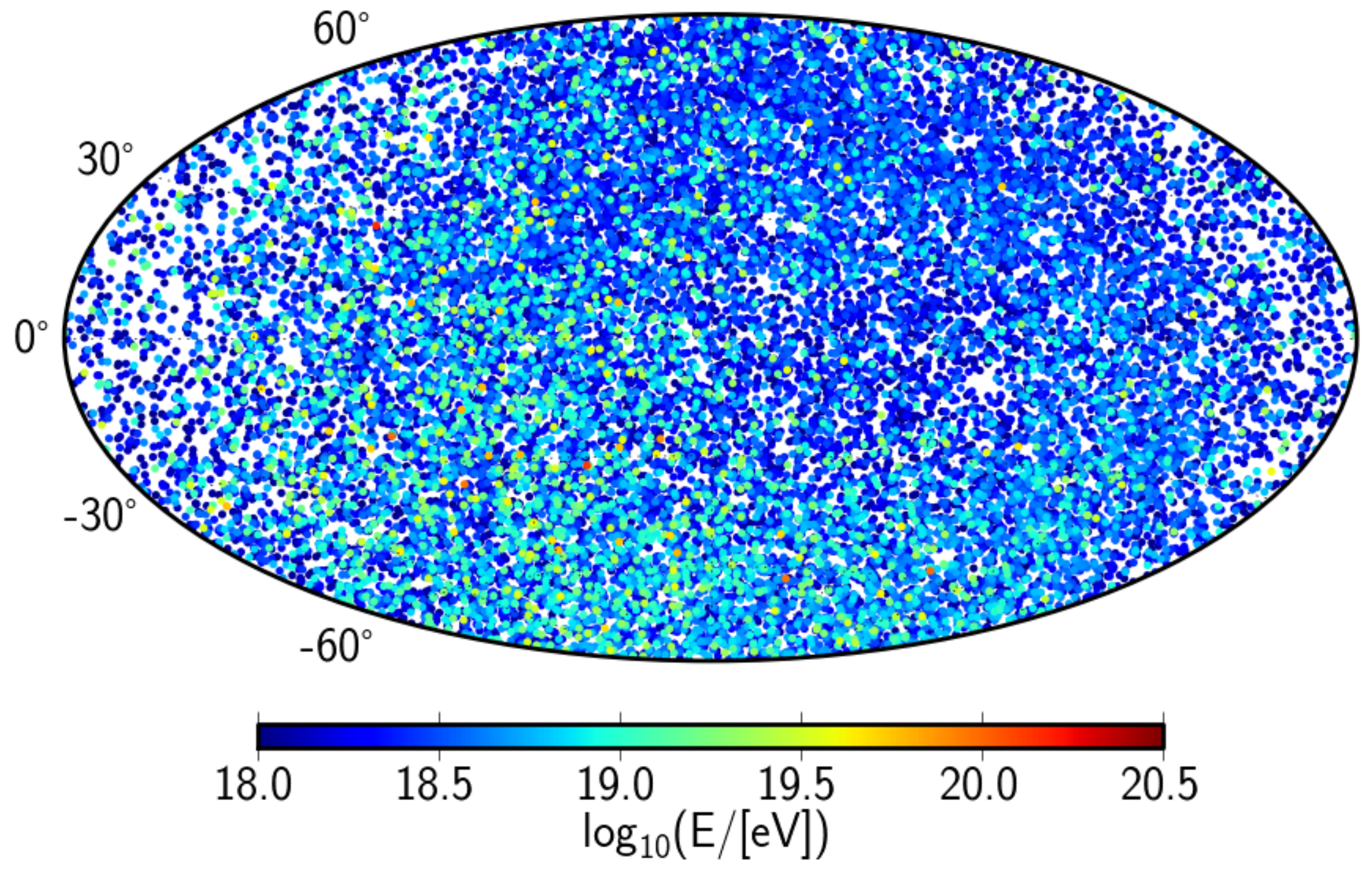}
        \label{fig:iron_gal}
    }
    \caption[Titel des Bildes]{In \subref{fig:spec} the spectra of UHECRs for injected protons (black) and iron (red) are shown. The solid lines are propagated in the Miniati LSS simulation, while the dashed lines show the universal spectrum for the respective species. In \subref{fig:defl} the mean deflection (angle between initial and observed momentum) vs.\ observed energy is shown. Again black denotes proton and red iron injection. The solid lines show the mean deflection in the Miniati EGMF, while the dashed lines show the deflection in the JF12 GMF. The thin black line denotes a mean deflection of $90^\circ$ which is the expectation for the complete decorrelation of initial and observed momentum.
The other figures show the arrival directions of injected proton and iron UHECRs at the border of the Galaxy \subref{fig:proton_xgal}, \subref{fig:iron_xgal} and at earth \subref{fig:proton_gal}, \subref{fig:iron_gal}. The color code denotes the observed energies of the particles.}
\end{figure*}

After extragalactic propagation, the deflections of the JF12 GMF model are applied to the arrival directions. Fig.\ \ref{fig:spec} shows the observed spectra. The spectra of initial protons and initial iron nuclei are drawn in solid black and red, respectively. For comparison, the universal spectra are plotted as dashed lines. Fig.\ \ref{fig:defl} shows the mean deflection, defined by the angle between initial and observed direction. As before red denotes iron and black proton injection. The solid lines show the mean deflections during extragalactic propagation, the dashed lines show the deflections in the GMF. Fig.\ \ref{fig:proton_xgal} and \ref{fig:iron_xgal} show the arrival directions for proton and iron injection before deflection in the GMF, while Fig.\ \ref{fig:proton_gal} and \ref{fig:iron_gal} show the arrival directions after deflections in the GMF.

From these figures can be seen that the magnetic field has a strong influence on the spectrum, especially for nuclei injection. This deviation from the universal spectrum can be explained by the strong magnetic field around the large scale structure in the Miniati simulation. As can be seen in Fig.\ \ref{fig:defl} the UHECRs lose their directional information almost completely during extragalactic propagation. Since the observer is placed in a low B-Field region, the containment of UHECRs in the large scale structure can cause a significant anisotropy in the arrival directions. In particular in the case of protons the GMF is not strong enough to completely isotropize the arrival directions, as can be seen in Fig.\ \ref{fig:proton_gal}.

\section{Benchmark tests}
To demonstrate the code performance two typical use cases are considered and the runtime compared to that of CRPropa 2.0.
The tests are performed on an i5-3317U CPU at 1.7 GHz.
As the runtime is highly dependent on the simulation settings the following values should be considered as estimates.\\
The first case is a 1D simulation with a uniform source distribution emitting a mixed composition of protons, helium-, nitrogen- and iron-nuclei with a log-flat energy spectrum between 1-1000\,EeV. CRPropa 3.0 runs in 4.5\,ms per injected particle, compared to 8.7\,ms using CRPropa 2.0.\\
The second case considers the 3D propagation of protons between 1-1000\,EeV in a turbulent magnetic field of 1\,nG $B_{\rm rms}$ strength over a distance of 1\,Gpc, while neglecting energy loss processes.
Here, CRPropa 3.0 runs in 7.5\,ms per trajectory, compared to 3.3\,ms using CRPropa 2.0.
The lower performance is due to the unoptimized compatibility mode for CRPropa 2.0 steering cards.
Running the same simulation with Python-steering, CRPropa 3.0 achieves 1.0\,ms per trajectory.\\
CRPropa 3.0 can make use of parallelization, thereby further reducing runtimes.
In a typical simulation the resulting speedup scales well up to 8 threads.
Thus, on a computing cluster with a standard 2\,GB RAM per core, a CRPropa 3.0 simulation can efficiently run on 8 cores in parallel, providing 16\,GB RAM for simulation data.
\section{Summary}
In this contribution we have introduced the public cosmic ray propagation code CRPropa 3.0. We
summarized its structure and highlighted new features, notably the possibility to take into account
cosmological redshift in 3-dimensional simulations, and the deflection in galactic magnetic fields.
We applied the code to a scenario in which either protons or iron nuclei are injected at sources following
the large scale structure from a simulation by Miniati et al.\ \cite{Sigl:2004} which also includes
cosmological magnetic field. Spectra, deflection angles and sky plots were shown for this scenario.
Finally, we performed some benchmark tests.\\
More information on CRPropa can be found on https://crpropa.desy.de.

\end{document}